# The running of the bare coupling in SU(N) gauge theories

**Chris Allton, Aurora Trivini**
*Dept of Physics, Swansea University, SA2 8PP, UK*
*E-mail:* `c.allton@swansea.ac.uk,pyat@swansea.ac.uk`

**Michael Teper***
*Dept of Physics, University of Oxford, OX1 3NP, UK*
*E-mail:* `m.teper1@physics.ox.ac.uk`

For $N \geq 5$ there is a first order bulk transition that cleanly separates the strong and weak coupling regimes of SU($N$) lattice gauge theories with the plaquette action. We find that in this case the calculated string tension can be readily fitted throughout the weak coupling region by a standard 3-loop perturbative expression modified by lattice spacing corrections of the expected form. While our fits demand the presence of the latter, they are not constraining enough to tell us which of the various bare coupling schemes is a 'good' one, in the sense that terms in the $\beta$-function beyond 3-loops are indeed negligible (in the relevant range of scales). To resolve this ambiguity we work in SU(3), using the Schrodinger Functional coupling scheme as a benchmark, and find that the Parisi mean-field improved coupling scheme matches it very well. Using the latter scheme, we have fitted the values of the string tension $a^2\sigma$ that have been calculated for $2 \geq N \geq 8$, to obtain $\Lambda_{\overline{MS}}/\sqrt{\sigma} = 0.503(2)(40) + 0.33(3)(3)/N^2$ for $N \geq 3$, where the first error is statistical and the second is our estimate of the systematic error from all sources.

---

*Speaker.

## 1. Introduction

Consider SU(*N*) lattice gauge theories with the standard plaquette action:

$$Z = \int \prod_l dU_l \exp\left\{-\beta \sum_p \left\{1 - \frac{1}{N}\text{ReTr}U_p\right\}\right\} \quad (1.1)$$

where $U_p$ is the ordered product of the SU(*N*) matrices around the boundary of the plaquette *p*. The parameter $\beta$ is the inverse bare coupling, and this defines a running coupling on the scale *a* in what one can call the 'Lattice' coupling scheme:

$$\beta = \frac{2N}{g_L^2(a)}. \quad (1.2)$$

It would be convenient to be able to determine *a* in units of a physical quantity, say the string tension $\sigma$, from the value of $g_L^2(a)$ using a weak coupling expansion of the form:

$$a\sqrt{\sigma}(a) \simeq \frac{\sqrt{\sigma}(0)}{\Lambda_L}\left(1 + ca^2\sigma + O(a^4)\right) F_{PT}(g_L^2(a)) \quad (1.3)$$

where $F_{PT}(g_L^2(a))$ is obtained by integrating the continuum $\beta$-function at some (practical) order in perturbation theory. The additional factor containing an $O(a^2)$ correction with coefficient $c \sim O(1)$ must be there [1] since if we were to use some other physical quantity $\mu'$ in place of $\mu \equiv \sqrt{\sigma}$ we would in general have

$$\frac{\mu'(a)}{\mu(a)} = \frac{\mu'(0)}{\mu(0)}\left(1 + c'a^2\mu^2 + O(a^4)\right), \quad (1.4)$$

with $c' \sim O(1)$, not to mention any $O(a^2)$ corrections from the $\beta$-function on the lattice.

There are two well-known problems with implementing this:

• $g_L^2$ is a poor expansion parameter, as indicated by

$$\frac{\Lambda_{\overline{MS}}}{\Lambda_L} = 38.853 \exp\left\{-\frac{3\pi^2}{11N^2}\right\}, \quad (1.5)$$

which implies that the *L* scheme will have large higher order terms in the $\beta$-function (assuming that the $\overline{MS}$ scheme is a 'good' one and does not);

• it is not clear at what $\beta$ we should expect such a weak coupling expansion to begin to work well, since SU(3) has a smooth strong-to-weak coupling crossover where

$$\text{powers in } \beta \quad \rightarrow \quad \text{powers in } \frac{1}{\beta}, \quad (1.6)$$

and this makes it hard to evaluate the relative merit of an 'improvement' to the lattice-scheme from an apparent success in fitting a wider range of bare couplings.

In this talk we describe the following strategy to resolve these two obstacles. First we use the fact that for SU($N \geq 5$) there is a first order 'bulk' transition [2], that separates the weak and strong coupling ranges, thus removing the ambiguity of where one might expect a weak coupling expansion to be applicable. (Just like the Gross-Witten transition [3] in $D = 2$.) While this enables us to quantify the importance of retaining $O(a^2)$ lattice corrections, it does not enable us to usefully discriminate between various bare coupling schemes which lead to quite different values for $\Lambda_{\overline{MS}}/\sqrt{\sigma}$. Presumably some have large higher order corrections in their $\beta$-function and so are 'bad'. To determine which of the schemes are 'good' ones we return to SU(3) and make use of the accurate calculation of the running coupling in the 'Schrodinger functional' (SF) scheme, that covers an energy range comparable to that of experiment, i.e. up to $\sim M_Z$, and with appreciably smaller errors [4]. We shall use this scheme to obtain, from the values of $a/r_0$ calculated in [5] the continuum value of $r_0 \Lambda_{SF}$ and hence of $r_0 \Lambda_{\overline{MS}}$. We compare this to what one obtains with various improved bare coupling extrapolations, and find that the original Parisi mean-field improved scheme [6] closely matches the SF result. We simultaneously perform a comparison with the SF scheme that does not involve the calculation of a physical quantity and therefore can be carried out to much weaker coupling. This also points to the 'goodness' of the mean-field scheme. Motivated by this we use the latter scheme for $N \neq 3$ to obtain continuum values for $\Lambda_{\overline{MS}}/\sqrt{\sigma}$ for all $N$, and in particular for $N \to \infty$.

In this talk we present a brief summary of our work: details, including estimates of the various systematic errors, will be published elsewhere [7].

## 2. Lessons from larger $N$

In Fig.1 we see the bulk transition, and its large metastability region, for SU(8).

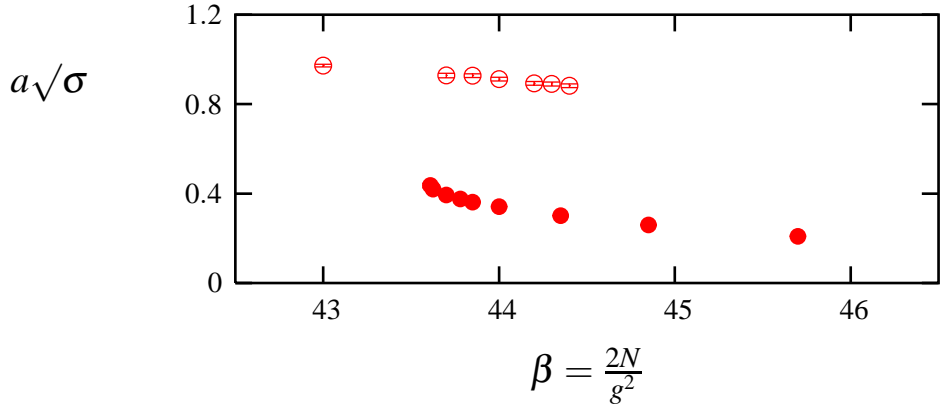

**Figure 1:** The SU(8) string tension versus the inverse lattice coupling, including the region of the first order 'bulk' transition between strong and weak coupling. Values ∘ are obtained coming from strong coupling, while the values • are obtained coming from weak coupling.

In Fig.2 we show a fit to the weak coupling branch, all the way to the extreme metastability edge, using

$$a\sqrt{\sigma}(a) = \frac{\sqrt{\sigma}(0)}{\Lambda_I}\left(1+ca^2\sigma\right)e^{-\frac{1}{2\beta_0 g_I^2}}\left(\frac{\beta_1}{\beta_0^2}+\frac{1}{\beta_0 g_I^2}\right)^{\frac{\beta_1}{2\beta_0^2}} e^{-\frac{\beta_1^I}{2\beta_0^2}g_I^2} \qquad (2.1)$$

where the scheme being used is the Parisi Mean Field Improved coupling [6]

$$\frac{1}{g_I^2} = \frac{1}{g_L^2}\langle\frac{1}{N}\mathrm{Tr}U_p\rangle \qquad (2.2)$$

where $U_p$ is the plaquette variable. In eqn(2.1) the terms that involve only $\beta_0$ and $\beta_1$ constitute the exact 2-loop continuum result. (That is to say, it is the exact result when $\beta_{j\geq 2} = 0$.) We present the 3-loop contribution as a power series in $g^2$. We note that although the coefficient $c$ is actually a power series in $g_I^2$, within our accuracy it suffices to treat it as a constant.

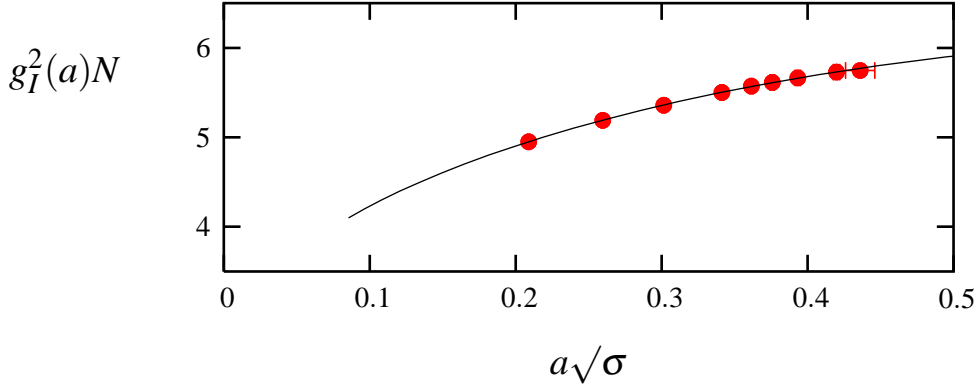

**Figure 2:** The 't Hooft coupling, defined from the mean-field improved lattice bare coupling as a function of the scale $a$ in SU(8). Shown is the 3-loop perturbative running modified by a $O(a^2)$ lattice correction.

The fit to SU(8) has $c = 1.18 \pm 0.04$ confirming the need for $O(a^2)$ corrections with coefficients of $O(1)$. However if we vary the perturbative coupling scheme we find that the range and accuracy of our calculations does not discriminate usefully between them.

Comparing the values of $g_I^2(a)N$ for various $N$ at fixed $a\sqrt{\sigma}$, shows good evidence for a large-$N$ $\beta$-function with very small corrections except at coarse lattice spacings. Thus it makes sense to take what we learn in SU(8) as a basis for treating other $N$, in particular SU(3). Performing fits with eqn(2.1) in SU(3) one sees in Fig 3 that these are only acceptable for $\beta \geq 5.9$, i.e. $a\sqrt{\sigma} \leq 0.25$, in contrast to the range $a\sqrt{\sigma} \leq 0.42$ for SU(8). For SU(2) the range is even more limited, i.e. $a\sqrt{\sigma} \leq 0.18$. This shows explicitly how the smoothening of the strong to weak coupling transition means that one has to go to much smaller values of $a$ to be able to use weak coupling expansions.

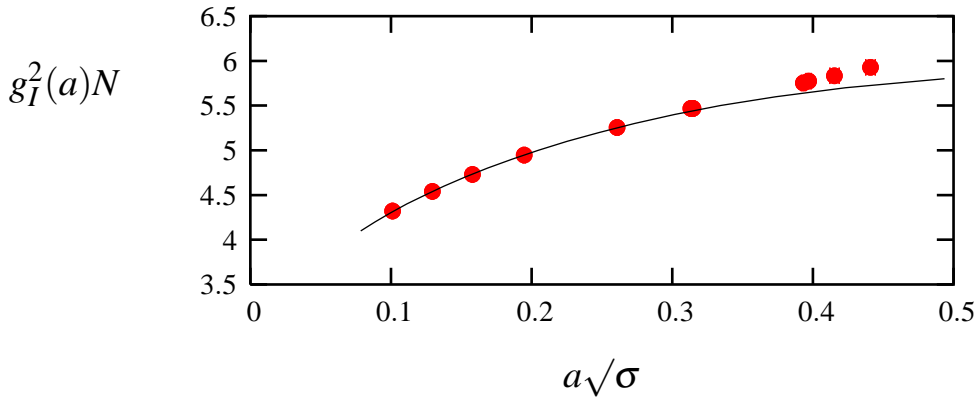

**Figure 3:** The 't Hooft coupling, defined from the mean-field improved lattice bare coupling as a function of the scale $a$ in SU(3). Shown is the 3-loop perturbative running modified by a $O(a^2)$ lattice correction.

## 3. Choosing a good coupling scheme

To choose a good bare coupling scheme $s$, we calculate $\Lambda_s/\mu$ and hence $\Lambda_{\overline{MS}}/\mu$, within various such schemes (for some physical mass $\mu$) and find which scheme produces values that agree with what we obtain using a 'reliable' lattice coupling scheme. For the latter we take the Schrodinger functional scheme of the Alpha Collaboration which for SU(3) [4] covers a range of energy scales comparable to that covered by experimental measurements, and does so with greater precision. (Compare Fig.4 of [4] with Fig.10 of [8].) The coupling $g_{SF}^2$ has been calculated for a wide variety of values of $\beta$ on scales $la(\beta)$ where typically $l = 6$ to $12$. We then take the calculated values of $r_0/a$ in [5] and interpolate these to the values of $\beta$ at which $g_{SF}^2(la)$ has been calculated. (Interpolating, unlike extrapolating, is a well controlled process.) We then fit using

$$\frac{la}{r_0(a)} = \frac{1}{r_0 \Lambda_{SF}} \left(1 + c_r^{SF} \frac{a^2}{r_0^2} + d_r^{SF} \frac{1}{l^p}\right)$$
$$\times e^{-\frac{1}{2\beta_0 g_{SF}^2(la)}} \left(\frac{\beta_1}{\beta_0^2} + \frac{1}{\beta_0 g_{SF}^2(la)}\right)^{\frac{\beta_1}{2\beta_0^2}} e^{-\frac{\beta_2^{SF}}{2\beta_0^2} g_{SF}^2(la)}. \tag{3.1}$$

Here there are two lattice spacing corrections. The usual $O(a^2)$ term arises from corrections to $r_0(a)$ etc. while the $O(1/l^p)$ term arises from lattice corrections to $g_{SF}^2(la)$ on the scale $l \times a$. We perform fits with both $p=1$ and $p=2$ taking the difference as part of our estimate of the systematic error. We obtain

$$\frac{1}{r_0 \Lambda_{SF}} = 3.2(1) \longrightarrow r_0 \Lambda_{\overline{MS}} = 0.640(20) \tag{3.2}$$

We now repeat this calculation using several lattice bare coupling schemes in fits of the form in eqn(2.1) but with $a\sqrt{\sigma(a)}$ replaced by $a/r_0(a)$. For the Parisi mean field improved coupling we find

$$\frac{1}{r_0 \Lambda_I} = 4.22(2) \longrightarrow r_0 \Lambda_{\overline{MS}} = 0.625(3) \tag{3.3}$$

which is consistent with the value in eqn(3.2), demonstrating that this coupling scheme is a reasonably good one. By contrast if we use a fit with the unadorned lattice bare coupling, $g_L^2(a)$, we find $r_0 \Lambda_{\overline{MS}} = 0.541(3)$ which demonstrates that this is not a good coupling scheme. We can also modify the mean field coupling scheme by replacing the true value of the plaquette in eqn(2.2) with its perturbative expansion up to $j$-loops. We call this coupling scheme $g_{I_j}^2$. These $I_j$ schemes will all have the same $\Lambda$ parameter (since this depends on a 1-loop relation) however we find they work much less well than the $I$ scheme. For example, the 1-loop improved coupling, $I_1$, gives a fit leading to $r_0 \Lambda_{\overline{MS}} = 0.448(2)$ – even worse than the bare lattice scheme!

There is also a way to compare schemes directly, without needing an extra physical quantity like $a/r_0(a)$. This has the advantage that one can perform comparisons deeper into weak coupling. For a scheme $s$ define the 3-loop perturbative factor

$$F_3^s[g_s^2] = e^{-\frac{1}{2\beta_0 g_s^2}} \left(\frac{\beta_1}{\beta_0^2} + \frac{1}{\beta_0 g_s^2}\right)^{\frac{\beta_1}{2\beta_0^2}} e^{-\frac{\beta_2^s}{2\beta_0^2} g_s^2}. \tag{3.4}$$

Now we expect for the SF scheme

$$la \Lambda_{SF} = \left\{1 + \frac{c_1}{l^p}\right\} F_3^{SF}[g_{SF}^2(al)] \tag{3.5}$$

and for a lattice improved scheme

$$a \Lambda_I = \left\{1 + c'a^2\right\} F_3^I[g_I^2(a)] \tag{3.6}$$

up to the various higher order corrections. If we now replace the $a^2$ on the RHS of eqn(3.6) by the expression for $a$ in eqn(3.5), and if we then take the ratio of the two equations, we obtain

$$\frac{\Lambda_{SF}}{\Lambda_I} = c_0 = \frac{1}{l} \frac{F_3^{SF}[g_{SF}^2(al)]}{F_3^I[g_I^2(a)]} \frac{\{1 + \frac{c_1}{l^p}\}}{\left\{1 + c_2 \frac{1}{l^2} \{1 + \frac{c_1}{l^p}\}^2 \{F_3^{SF}[g_{SF}^2(al)]\}^2\right\}}. \quad (3.7)$$

We can now perform a fit for the constants $c_0$, $c_1$ and $c_2$ over $\beta$ ranges further and further into weak coupling, and see how rapidly $c_0$ approaches the known value of $\Lambda_{SF}/\Lambda_I$. In Fig. 4 we show a comparison for three schemes. Again we see that the Parisi scheme works well – and much better than the other schemes shown.

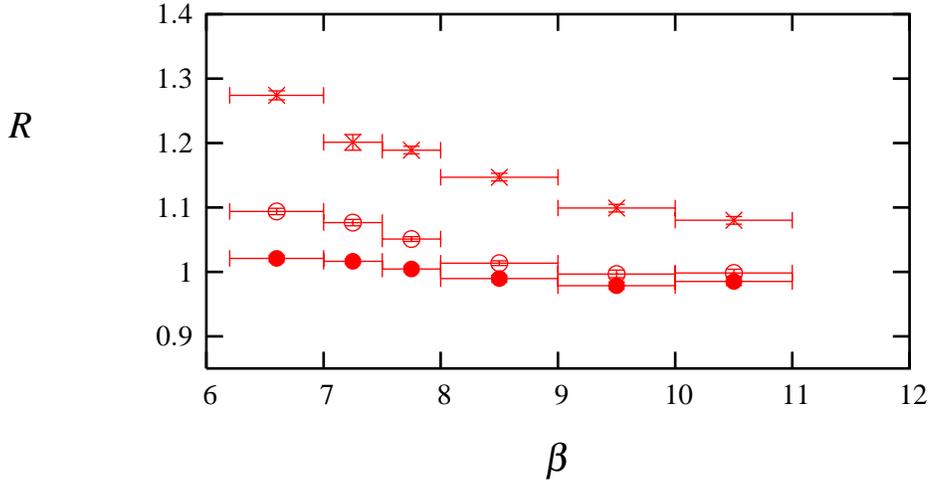

**Figure 4:** Calculated values of $\Lambda_{SF}/\Lambda_s$ for the $s = I$, •, $s = I_3$, ∘, and the $s = L$, ×, lattice bare coupling schemes, all normalised to the known theoretical values. Horizontal errors indicate the range of $\beta$ values used in each fit.

## 4. Conclusions

Taking advantage of the fact that large $N$ lattice gauge theories have a well-defined weak coupling branch, we saw quite explicitly that $O(a^2)$ lattice spacing corrections are indeed important for transmuting the value of the bare lattice coupling into a value of the lattice spacing in 'physical' units [1].

We have also learned that the Parisi mean-field improvement scheme [6] for the bare coupling is in fact a reasonably good one. This we did by comparing it to the Schrodinger Functional scheme which we used as our benchmark. Obviously it will not be unique in this respect, and one could pursue this programme further. One cautionary remark: our benchmark *SF* coupling is defined in a finite volume, and one needs to understand the implications for this of the finite volume phase transitions at $N = \infty$ [9] that will lead to cross-overs at finite $N$.

We can use fits of the form eqn(2.1) to extract values of $\Lambda_I/\sqrt{\sigma}$ and hence $\Lambda_{\overline{MS}}/\sqrt{\sigma}$ for all $N$. Doing so, in Fig. 5, we find that these values can be fitted with a modest $O(1/N^2)$ correction

$$\frac{\Lambda_{\overline{MS}}}{\sqrt{\sigma}} = 0.503(2)(40) + \frac{0.33(3)(3)}{N^2} \quad ; \quad N \geq 3 \quad (4.1)$$

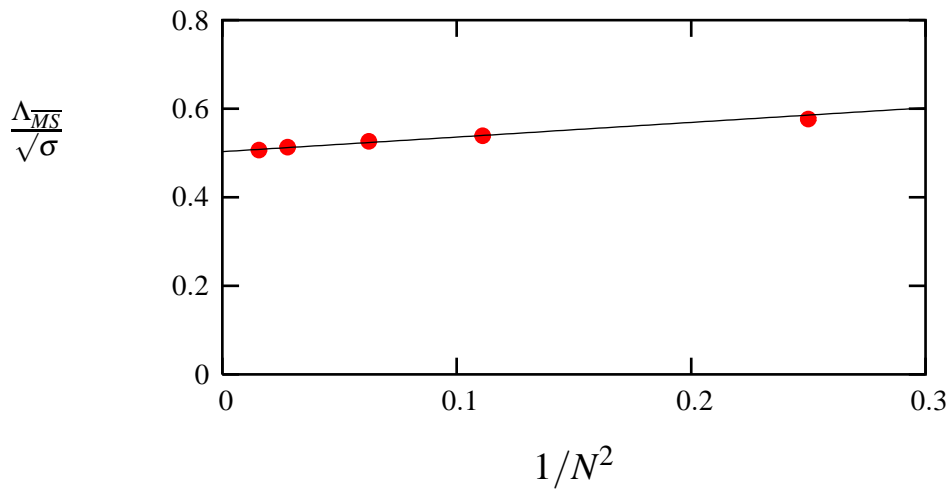

**Figure 5:** Calculated values of $\Lambda_{\overline{MS}}/\sqrt{\sigma}$ versus $1/N^2$ with a linear extrapolation to $N = \infty$ shown.

(We choose to exclude SU(2) from the fit, because of the difficulty in identifying a region where a weak coupling expansion is valid, but our fit does agree, when extrapolated to $N = 2$, with the value naively obtained there.) Here the first error is statistical and the second much larger error is expected to provide a bound on the systematic error from all sources.